\documentclass[twocolumn]{aastex63}

\hypersetup{linkcolor=blue,citecolor=blue,filecolor=cyan,urlcolor=blue}

\usepackage{CJK}

\usepackage{longtable}
\usepackage{lipsum} 
\usepackage{multirow}

\usepackage{natbib}
\usepackage{graphicx}
\usepackage{bm}
\usepackage{amsmath}
\usepackage{amsfonts}
\usepackage{amssymb}
\usepackage{xcolor}
\usepackage{mathrsfs}
\usepackage[caption=false]{subfig}
\usepackage{CJK}
\usepackage{ulem}

\shorttitle{NGC\,6334S}
\shortauthors{Li et al.}
\usepackage{CJK}

\graphicspath{{./}{figures/}}
\usepackage{xspace}
\newcommand{\um}{{$\mu{}$m}\xspace}
\def\arcsec{$^{\prime\prime}$}
\def\Mo{$M_{\odot}$\xspace}
\def\kms{km s$^{-1}$\xspace}

\def\1{\uppercase\expandafter{\romannumeral1}}
\def\2{\uppercase\expandafter{\romannumeral2}}

\newcommand{\x}{\textcolor{red}{XX}}

\begin{document}

\title{A Low-mass Cold and Quiescent Core Population in a Massive Star Protocluster}

\correspondingauthor{Shanghuo Li}
\email{shanghuo.li@gmail.com; ktkim@kasi.re.kr}

\author[0000-0003-1275-5251]{Shanghuo Li }
\affiliation{Korea Astronomy and Space Science Institute, 776 Daedeokdae-ro, Yuseong-gu, Daejeon 34055, Republic of Korea}
\affiliation{Shanghai Astronomical Observatory, Chinese Academy of Sciences, 80 Nandan Road, Shanghai 200030, People's Republic of China}
\affiliation{Center for Astrophysics $|$ Harvard \& Smithsonian, 60 Garden Street, Cambridge, MA 02138, USA}
\affiliation{University of Chinese Academy of Sciences, 19A Yuquanlu, Beijing 100049, People's Republic of China}

\author{Xing Lu}
\affiliation{National Astronomical Observatory of Japan, National Institutes of Natural Sciences, 2-21-1 Osawa, Mitaka, Tokyo 181-8588, Japan}

\author{Qizhou Zhang}
\affiliation{Center for Astrophysics $|$ Harvard \& Smithsonian, 60 Garden Street, Cambridge, MA 02138, USA}

\author{Chang-Won Lee}
\affil{Korea Astronomy and Space Science Institute, 776 Daedeokdae-ro, Yuseong-gu, Daejeon 34055, Republic of Korea}
\affil{University of Science and Technology, 217 Gajeong-ro, Yuseong-gu, Daejeon 34113, Republic of Korea}

\author{Patricio Sanhueza}
\affiliation{National Astronomical Observatory of Japan, National Institutes of Natural Sciences, 2-21-1 Osawa, Mitaka, Tokyo 181-8588, Japan}
\affiliation{Department of Astronomical Science, SOKENDAI (The Graduate University for Advanced Studies), 2-21-1 Osawa, Mitaka, Tokyo 181-8588, Japan}

\author{Henrik Beuther}
\affiliation{Max Planck Institute for Astronomy, Konigstuhl 17, 69117 Heidelberg, Germany}

\author{Izaskun, Jim\'enez-Serra}
\affiliation{Centro de Astrobiolog\'ia (CSIC-INTA), Ctra. de Torrej\'on a Ajalvir, Km. 4, Torrej\'on de Ardoz, 28850 Madrid, Spain}

\author{Keping Qiu}
\affiliation{School of Astronomy and Space Science, Nanjing University, 163 Xianlin Avenue, Nanjing 210023, People's Republic of China}

\author{Aina Palau}
\affiliation{Instituto de Radioastronom\'ia y Astrof\'isica, Universidad Nacional Aut\'onoma de M\'exico, P.O. Box 3-72, 58090, Morelia, Michoac\'an, M\'exico}

\author{Siyi Feng}
\affiliation{National Astronomical Observatories, Chinese Academy of Sciences, Beijing 100101, People's Republic of China}
\affiliation{Academia Sinica Institute of Astronomy and Astrophysics, No.\ 1, Section 4, Roosevelt Road, Taipei 10617, Taiwan, Republic of China}
\affiliation{National Astronomical Observatory of Japan, 2-21-1 Osawa, Mitaka, Tokyo, 181-8588, Japan}

\author{Thushara Pillai}
\affiliation{Institute for Astrophysical Research, Boston University, 725 Commonwealth Ave, Boston, MA 02215, USA}

\author{Kee-Tae Kim}
\affil{Korea Astronomy and Space Science Institute, 776 Daedeokdae-ro, Yuseong-gu, Daejeon 34055, Republic of Korea}
\affil{University of Science and Technology, 217 Gajeong-ro, Yuseong-gu, Daejeon 34113, Republic of Korea}

\author{Hong-Li Liu}
\affiliation{Department of Astronomy, Yunnan University, Kunming, 650091, People's Republic of China}

\author{Josep Miquel. Girart}
\affiliation{Institut de Ci\`encies de l'Espai (IEEC-CSIC), Campus UAB, Carrer de Can Magrans s/n, 08193 Cerdanyola del Vall\`es, Catalonia, Spain}

\author{Tie Liu}
\affiliation{Shanghai Astronomical Observatory, Chinese Academy of Sciences, 80 Nandan Road, Shanghai 200030, People's Republic of China}

\author{Junzhi Wang}
\affil{Shanghai Astronomical Observatory, Chinese Academy of Sciences, 80 Nandan Road, Shanghai 200030, People's Republic of China}

\author{Ke Wang}
\affil{Kavli Institute for Astronomy and Astrophysics, Peking University, 5 Yiheyuan Road, Haidian District, Beijing 100871, People's Republic of China}

\author{Hauyu Baobab Liu}
\affiliation{Academia Sinica Institute of Astronomy and Astrophysics, 11F of AS/NTU Astronomy-Mathematics Building, No.1, Sec. 4, Roosevelt Rd, Taipei 10617, Taiwan, Republic of China}

\author{Howard A. Smith}
\affiliation{Center for Astrophysics $|$ Harvard \& Smithsonian, 60 Garden Street, Cambridge, MA 02138, USA}

\author{Di Li}
\affiliation{National Astronomical Observatories, Chinese Academy of Sciences, Beijing 100101, People's Republic of China}
\affiliation{NAOC-UKZN Computational Astrophysics Centre, University of KwaZulu-Natal, Durban 4000, South Africa}

\author{Jeong-Eun Lee}
\affiliation{School of Space Research, Kyung Hee University, 1732, Deogyeong-Daero, Giheung-gu, Yongin-shi, Gyunggi-do 17104,  Republic of Korea}

\author{Fei Li} 
\affiliation{School of Astronomy and Space Science, Nanjing University, 163 Xianlin Avenue, Nanjing 210023, People's Republic of China}

\author{Juan Li}
\affil{Shanghai Astronomical Observatory, Chinese Academy of Sciences, 80 Nandan Road, Shanghai 200030, People's Republic of China}

\author{Shinyoung Kim}
\affil{Korea Astronomy and Space Science Institute, 776 Daedeokdae-ro, Yuseong-gu, Daejeon 34055, Republic of Korea}
\affil{University of Science and Technology, 217 Gajeong-ro, Yuseong-gu, Daejeon 34113, Republic of Korea}

\author{Nannan Yue}
\affil{Kavli Institute for Astronomy and Astrophysics, Peking University, 5 Yiheyuan Road, Haidian District, Beijing 100871, People's Republic of China}

\author{Shaye Strom}
\affiliation{Center for Astrophysics $|$ Harvard \& Smithsonian, 60 Garden Street, Cambridge, MA 02138, USA}




\begin{abstract}
Pre-stellar cores represent the initial conditions of star formation. 
Although these initial conditions in nearby low-mass star-forming 
regions have been investigated in detail, such initial conditions remain 
vastly unexplored for massive star-forming regions. 
We report the detection of a cluster of  low-mass starless 
and pre-stellar core candidates in a massive star protocluster forming 
cloud, NGC\,6334S. With the ALMA observations at a $\sim$0.02 pc 
spatial resolution, we identified 17 low-mass starless core candidates 
that do not show any evidence of protostellar activity. These 
candidates present small velocity dispersions, high fractional 
abundances of NH$_{2}$D, high NH$_{3}$ deuterium fractionations, 
and are completely dark in the infrared wavelengths from 3.6 up to 
70~$\mu$m. Turbulence is significantly dissipated and the gas kinematics 
are dominated by thermal motions toward these candidates. 
Nine out of the 17 cores are gravitationally bound, and therefore 
are identified as pre-stellar core candidates. The embedded cores 
of NGC\,6334S show a wide diversity in masses and evolutionary stages.

\end{abstract}


\section{Introduction} 
\label{sec:intro}
Stars form in the high-density parts of molecular clouds, 
i.e., dense cores.  A dense core prior to the protostellar 
phase is called a starless core, and a gravitationally 
bound/unstable starless core is referred to as a pre-stellar 
core \citep{1994MNRAS.268..276W}. 
The pre-stellar phase is considered as the starting point in 
the star formation process \citep{2007ARA&A..45..339B,
2012A&ARv..20...56C}. 
Pre-stellar cores can form single or multiple stellar systems 
under the combined effect of gravity, magnetic fields, and 
turbulence \citep{2015Natur.518..213P}. 
Understanding the properties of pre-stellar cores is critical to 
characterize the initial conditions of cluster formation. 
The pre-stellar cores in low-mass star-forming regions have 
been intensively studied toward nearby molecular clouds, 
e.g., Perseus, Ophiuchus, Chamaeleon, Serpens, and Taurus 
\citep{1999ApJ...526..788L,2001ApJS..136..703L,
2008ApJ...684.1240E,2010ApJ...718..306S,
2016ApJ...823..160D,2017ApJ...838..114K,2020ApJ...899...10T}. 
However, the studies of pre-stellar cores in massive cluster-forming 
regions are still limited by the low number of statistics available 
\citep{2018A&A...618L...5N,2019ApJ...886..130L,
2019ApJ...886..102S,2021ApJ...907L..15S}.

The massive infrared dark cloud NGC\,6334S (also known as IRDC 
G350.56+0.44) is located at the southwestern end of the NGC\,6334 
molecular cloud complex that is a nearby  
\citep[1.3 kpc;][]{2014ApJ...784..114C} young and massive 
`mini-starburst' star-forming region \citep{2013ApJ...778...96W}. 
With a mass of 1.3 $\times \, 10^{3} \, M_{\odot}$ 
\citep{2020ApJ...896..110L}, comparable to the clumps with 
embedded massive protostars and protoclusters in the complex, 
NGC\,6334S has the potential to form a cluster 
encompassing both low- ($<$ 2 \Mo) and high-mass ($>$ 8 \Mo) stars. 
NGC\,6334S provides an ideal laboratory to search for and study the 
early stages (e.g., pre-stellar phase) of star formation in the high-mass 
regime. 

In order to investigate massive star and cluster formation, we performed 
Atacama Large Millimeter/submillimeter Array (ALMA) 
and Karl G. Jansky Very Large Array (JVLA) observations of NGC\,6334S.  
A study of 49 continuum dense cores (hereafter continuum cores) 
revealed by the 3~mm wavelength continuum image was presented in 
\citet{2020ApJ...896..110L}, in which we reveal that the nonthermal 
motions are predominantly subsonic and transonic in both clump scale 
and continuum core scale. 
Here we use molecular lines to study the starless/pre-stellar cores.

\section{Observations} 
\label{sec:obs}
\subsection{ALMA Observations} 
We carried out a 55--pointing mosaic of NGC\,6334S using 
the ALMA 12 m array in 2017 March (ID: 2016.1.00951.S). 
Two 234.4 MHz width spectral windows with a spectral resolution 
of 61~kHz ($\sim$0.21 \kms at 86 GHz) were configured to cover 
the H$^{13}$CO$^{+}$ (1--0, 86.754 GHz) and ortho-NH$_{2}$D 
($1_{11}-1_{01}$ , 85.926 GHz) lines, respectively. Three additional 
1.875 GHz wide spectral windows at 88.5 GHz, 98.5 GHz, and 
100.3 GHz with a coarse spectral resolution (3.0 -- 3.3 \kms) 
were employed to obtain broad band continuum 
and rotational 
transitions of several molecular species (e.g., HCO$^{+}$ 1--0, 
HCN 1--0, CS 2--1, HNCO $4_{0,4}-3_{0,3}$, H$^{15}$NC 1--0, 
CH$_{3}$OH $5_{1,4}-4_{1,3}$, SO $2_{2}-1_{1}$, 
HC$_{3}$N 11--10).  The maximum recoverable scales (MRS) 
is $\sim$25\arcsec\ in the ALMA data. 
The details of the observations can be found 
in \citet{2020ApJ...896..110L}.

Data calibration and imaging were performed using CASA 4.7.0 
\citep{2007ASPC..376..127M}. 
We used Briggs' robust weighting of 0.5 to the visibilities for both 
the continuum and lines, which results in a synthesized beam of 
3.\arcsec6 $\times$ 2.\arcsec4 with a position angle (P.A.) of 
81$^{\circ}$ and  4.\arcsec1 $\times$ 2.\arcsec8 (P.A. = 83$^{\circ}$) 
for continuum and line images, respectively. 
The achieved 1$\sigma$ rms noise level is about  6 mJy~beam$^{-1}$ 
per 0.21 \kms\ for the line images and 30~$\mu$Jy~beam$^{-1}$ for 
the continuum image. 
All images shown in the paper are prior to primary beam correction, 
while all measured fluxes are corrected for the primary beam attenuation.

\subsection{JVLA Observations} 
We carried out a 4-pointing mosaic of the central region of 
NGC\,6334S using the JVLA C-configuration in 2014 August 
(ID:14A-241).
The NH$_{3}$ (1, 1) through (5, 5) metastable inversion transitions 
and H$_{2}$O maser were simultaneously covered in this observation. 
CASA versions 4.7.0 and 5.1.1 were used to perform data 
calibration and imaging. 
An elliptical Gaussian with a FWHM of 6\arcsec $\times$ 3\arcsec 
(P.A. = 0$^{\circ}$) was used to taper the visibilities, in order to 
increase the signal-to-noise ratio (S/N). 
This yields a synthesized beam size of about 
10\arcsec $\times$ 5\arcsec (P.A. = 26$^{\circ}$) with a 1$\sigma$ 
rms noise of 9 mJy beam$^{-1}$ per 0.2 \kms for the NH$_{3}$ lines. 
More details on the observations are presented in \cite{2020ApJ...896..110L}.

\begin{figure*}[!ht]
\centering
\includegraphics[scale=0.28]{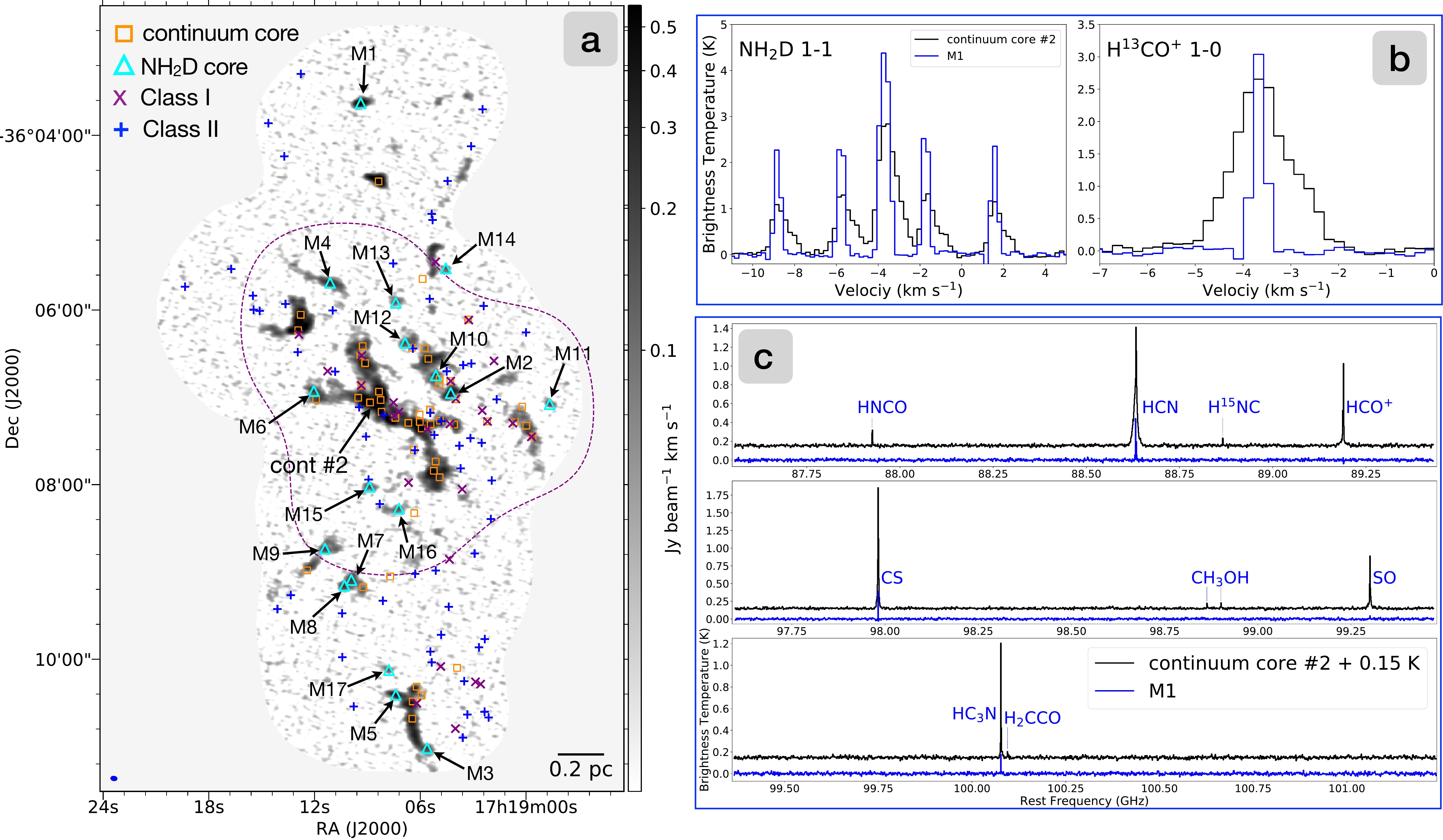}
\caption{ 
Panel  a: the gray scale in the background shows the NH$_{2}$D 
velocity-integrated intensity ($W_{\rm NH_{2}D}$) image. 
Cyan open triangles show the NH$_{2}$D  cores. 
Yellow open squares present 49 continuum cores identified by 
ALMA~3~mm continuum image \citep{2020ApJ...896..110L}. 
Purple cross `x' and blue plus `+' symbols are 25 Class I  and 58 Class II YSOs  
\citep{2013ApJ...778...96W}, respectively. 
The beam size of the NH$_{2}$D image is shown on the bottom 
left of the panel.  Dashed purple contour shows the area mosaicked 
with VLA. 
Panel b:  the core-averaged spectra of NH$_{2}$D ($1_{11}-1_{01}$) 
and H$^{13}$CO$^{+}$ (1-0) for continuum core \#2 (black solid line)  
and NH$_{2}$D core M1 (blue solid line). 
Panel c:  the core-averaged spectra of three wide (1.875 GHz) 
spectral windows for continuum core \#2 (black solid line) and 
NH$_{2}$D core M1 (blue solid line), respectively. 
}
 \label{fig:cont}
\end{figure*}

\begin{figure*}[!ht]
\centering
\includegraphics[scale=0.58]{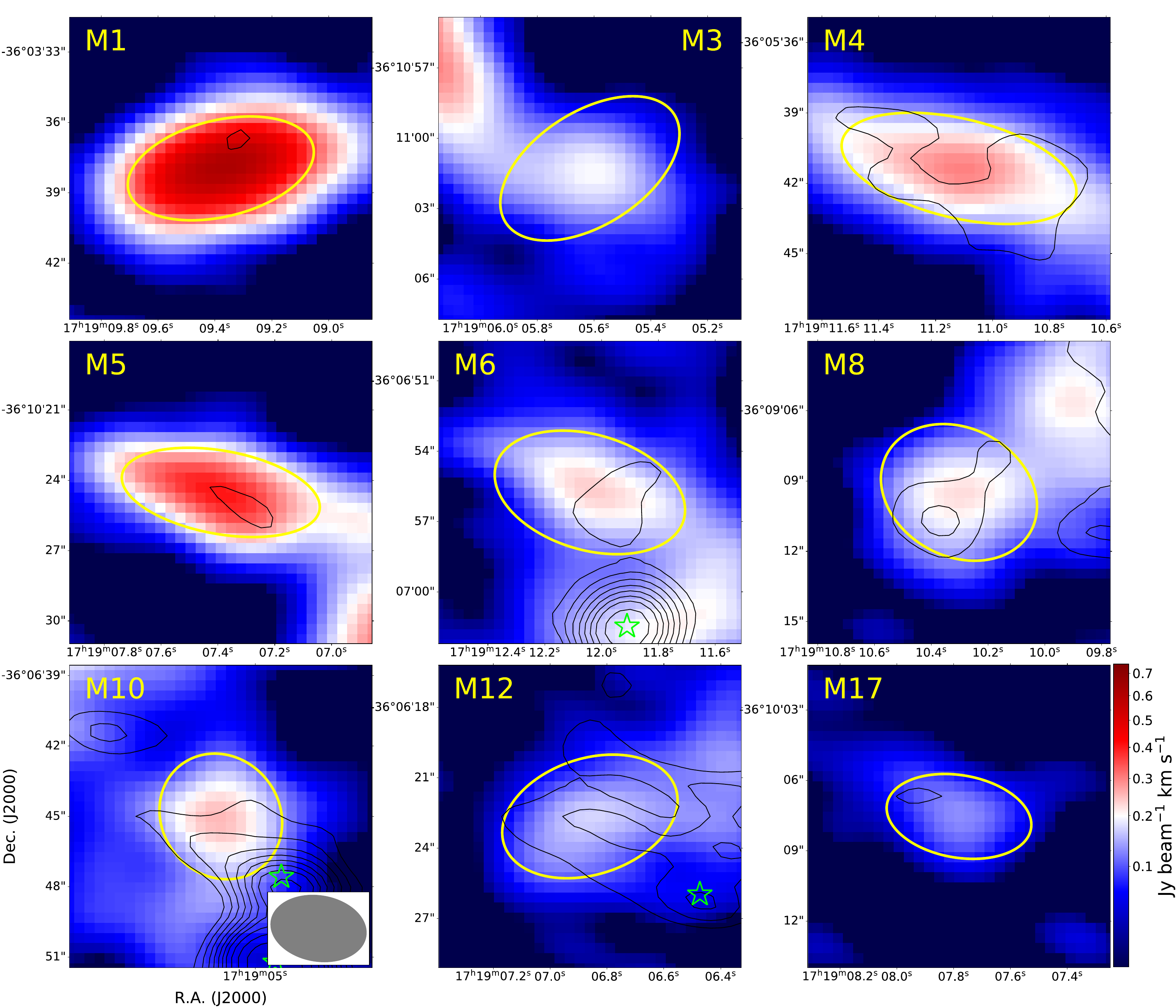}
\caption{
The 3~mm continuum image (black contours) overlaid on the 
velocity-integrated intensity of NH$_{2}$D for the candidates that 
are gravitationally bound with $\alpha_{\rm vir} <$ 2.  
The integrated velocity range is 
-11.5 and 5.6 \kms. Black contours are (5, 7, 9, 11, 15, 17, 
20, 24, 29)$\times \sigma$, where $\sigma$ = 30 $\mu$Jy~beam$^{-1}$ 
is the rms noise level of the continuum image. The yellow ellipses 
show the identified  NH$_{2}$D cores.  The green open stars 
indicate continuum cores. The beam size of NH$_{2}$D image 
is shown on the bottom right of M10 panel.  
}
 \label{fig:cores}
\end{figure*}

\section{Results and Analysis} 
\label{sec:results}
\subsection{Cold and Quiescent Cores} 
\label{sec:core}
NH$_{2}$D ($1_{11}-1_{01}$, critical density 
$n_{\rm cr} \sim$ 10$^{5}$ cm$^{-3}$) 
is a good tracer of cold and dense molecular gas 
\citep{2007A&A...470..221C,2013ApJ...773..123S}, which can survive 
in the gas phase in the dense interior region of pre-stellar cores 
\citep{2017A&A...600A..61H}.  In our ALMA data, 
the NH$_{2}$D ($1_{11}-1_{01}$) line 
emission is in general well correlated with the 3~mm continuum 
emission, but there are some exceptions; there are 17 bright 
compact structures in the NH$_{2}$D emission that are 
associated with weak (3$\sigma$ -- 11$\sigma$) or no 
continuum emission ($< 3\sigma$) at all (Figure~\ref{fig:cont} and 
\ref{fig:cores}),  and with no young stellar object (YSO) counterparts 
\citep[e.g., Class~\1/\2;][]{2013ApJ...778...96W}. 
There is no reliable Class~0 catalogue in this region.  
The details on the identification of NH$_{2}$D compact structures  
are summarized in Appendix~\ref{app:identification}. 
We refer to these NH$_{2}$D compact structures as NH$_{2}$D 
cores, naming those in M1, M2, M3 ... in order of descending 
NH$_{2}$D velocity-integrated intensity. 
They have diameters ranging from 0.018 to 0.04 pc (see 
Table~\ref{tab:nh2d}).

Among the NH$_{2}$D cores, M1 is the most prominent  
pre-stellar candidate with high signal-to-noise (S/N) ratio and a 
relatively isolated environment that eliminates the contamination 
from the outflows driven by the continuum cores. The following 
analysis and discussion will therefore use M1 as a showcase, 
while the physical parameters of the remaining NH$_{2}$D 
cores will only be included in statistical analyses.

Using the HCO$^{+}$, CS, HCN, SO, HNCO, and H$_{2}$O maser 
lines (see Figure~\ref{fig:cont}), we have searched for signatures 
of protostellar activity, such as  bipolar/monopolar/multipolar outflow 
activity  and H$_{2}$O maser emission.  All the NH$_{2}$D cores 
show no signs of protostellar activity.  
Although we currently cannot 
rule out the existence of unresolved or weak outflows due to the 
coarse spectral resolution and the lack of better low-mass outflow 
tracers such as CO, these NH$_{2}$D cores appear to be 
starless with the available evidence.  
In addition, the NH$_{2}$D 
cores do not show any appreciable ($< 3\sigma$) continuum emission 
in the infrared wavelengths from 3.6 up to 70 \um, and therefore they are 
completely dark in these infrared images. 
The upper limit on the internal luminosity ($L_{\rm int}$) of NH$_{2}$D 
cores is estimated to be $\sim$1.26 $L_{\odot}$ using 
\texttt{Herschel}/PACS 70 \um\ data \citep[][see Appendix~\ref{app:lum} 
for detailed derivation of internal luminosity]{2017A&A...602A..77T}.  
These results suggest that these NH$_{2}$D cores are cold and quiescent 
in terms of star-forming activity, though some undetected very faint 
embedded low-mass protostars cannot be fully ruled out 
\citep{2008ApJS..179..249D,2009ApJS..181..321E,2011ApJ...736...53O,
2017ApJ...840...69F}. 
An example of the comparison between NH$_{2}$D core (M1) 
and continuum core (\#2) is shown in Figure~\ref{fig:cont}, which 
includes all of the observed spectral windows. Continuum core \#2 is
a protostellar core as evidenced by the clear molecular outflows in the 
CS, HCN, and HCO$^{+}$ line emission, and it is one of the most 
chemically rich among the continuum cores. 
Nonetheless, only a few lines are detected toward continuum 
core \#2. This indicates that the continuum cores \#2 is at 
a stage prior to the hot cores/corinos.  We note that M1 shows even 
less chemical complexity than continuum core \#2, with fewer lines, 
molecular species, and narrower line profiles. This suggests that the 
evolutionary stage of the NH$_{2}$D cores are likely earlier than the 
continuum cores. 
On the other hand, we cannot completely rule out the possibility that 
some of continuum cores are relatively evolved pre-stellar cores, 
because of the lack of high sensitivity and high spectral resolution 
outflow tracers (e.g., CO, SiO) to distinguish the evolutionary stages 
\citep{2020ApJ...903..119L}. 
In this study, we will focus only on the NH$_{2}$D cores.

\begin{figure*}[!ht]
\centering
\includegraphics[scale=0.27]{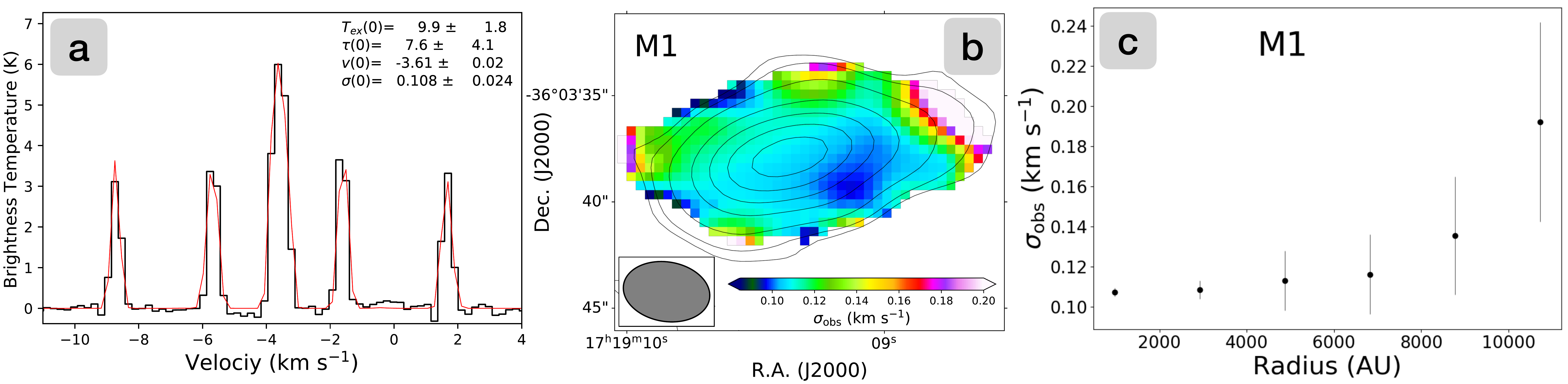}
\caption{Panel a: NH$_{2}$D spectrum (black curve) extracted 
from single pixels toward M1.  Red curve shows the best result from 
the hyperfine fitting that includes NH$_{2}$D  36 hyperfine 
components \citep{2011ascl.soft09001G,2016A&A...586L...4D}, 
with the best-fit parameters displayed in the panel.   
Panel  b: the NH$_{2}$D integrated intensity (black contours) 
overlaid on its $\sigma_{\rm obs}$ toward M1.  
The contours are (3, 5, 10, 20,  30, 40, 50)$\times \sigma$, 
with $\sigma$ = 0.011~Jy~beam$^{-1}$~km~s$^{-1}$. 
The beam size is shown on the lower left of the panel. 
Panel c: the annularly averaged observed velocity dispersions 
($\sigma_{\rm obs}$) as a functions of radial distance from the 
center (given in Table~\ref{tab:nh2d}) of M1.  
The error bars show the statistical standard deviation at each bin. 
 }
 \label{fig:sigma}
\end{figure*}

\subsection{Small Velocity Dispersions} 
\label{linewidth}
\begin{figure}[!ht]
\centering
\includegraphics[scale=0.311]{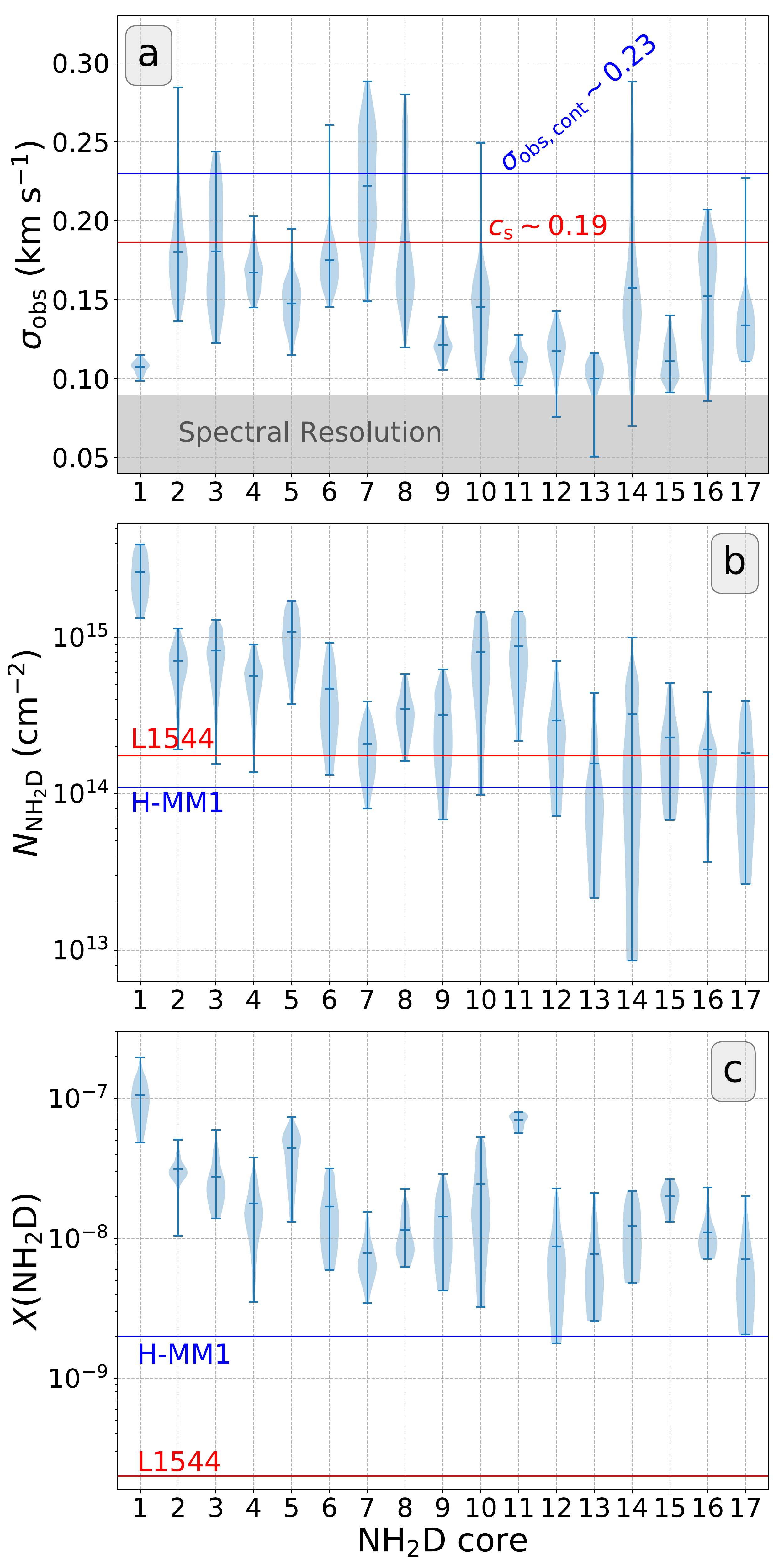}
\caption{
Panel a to c: violin plots of the $\sigma_{\rm obs}$, $N_{\rm NH_{2}D}$, 
and $\chi(\rm NH_{2}D)$ distributions for all NH$_{2}$D cores. 
The shape of each distribution shows the probability density of the data 
smoothed by a kernel density estimator. The blue bars from the top to 
bottom represent the maximum, mean, and minimum values, respectively. 
Panel a: the red solid line is the sound speed at a temperature of 10 K. 
The blue solid line is the mean observed velocity dispersion of the 
continuum cores, $\langle \sigma_{\rm obs, cont} \rangle$ = 0.23 $\pm$ 
0.08 \kms. 
The shaded gray region shows the spectral resolution limit of the 
NH$_{2}$D line. 
Panel b: the red and blue solid lines are observed $\rm N(NH_{2}D)$ 
in the pre-stellar core L1544 \citep[1.75 $\times\, 10^{14}$ cm$^{-2}$;][]{
2007A&A...470..221C} and 
Ophiuchus/H-MM1 \citep[1.1 $\times\, 10^{14}$ cm$^{-2}$;][]{
2017A&A...600A..61H}. 
Panel c: the blue and red solid lines are observed abundance 
$X(\rm NH_{2}D)$ in the pre-stellar cores Ophiuchus/H-MM1 
\citep[2 $\times\, 10^{-9}$,][]{2017A&A...600A..61H} and L1544 
\citep[2 $\times\, 10^{-10}$, ][]{2007A&A...470..221C}, respectively.   
}
 \label{fig:comb}
\end{figure}

The NH$_{2}$D line cube is fitted pixel by pixel, following the fitting 
processes described in \cite{2020ApJ...896..110L}. In the hyperfine fits, 
the uncertainties of the LSR velocities and line widths are small, while 
those of the excitation temperature and optical depth are relatively 
large (Figure~\ref{fig:sigma}). 
Figure~\ref{fig:comb} shows the observed velocity dispersions 
($\sigma_{\rm obs}$) distribution for each NH$_{2}$D core. All the 
NH$_{2}$D cores show small velocity dispersion 
($\langle \sigma_{\rm obs} \rangle \sim$ 0.16 \kms) 
and the majority of them are smaller than the sound speed, 
$c_{\rm s}$(10 K) = 0.19 \kms.  
The observed velocity dispersions may be regarded as upper limits, 
given the limited spectral resolution, the blending of velocity 
components along the line of sight and opacity broadening effects. 
For the NH$_{2}$D cores, the line widths of detected lines    
are smaller than those of the majority of continuum cores 
(see Figure~\ref{fig:cont} and \ref{fig:comb}).

Among all the NH$_{2}$D cores, M1 shows the strongest 
NH$_{2}$D emission and the smallest velocity dispersion. 
For M1, pixel by pixel fit to the NH$_{2}$D 
line cube yields a small mean velocity dispersion 
$\langle \sigma_{\rm obs} \rangle \approx$ 0.107 km~s$^{-1}$ 
(Figure~\ref{fig:comb}), corresponding to an intrinsic observed 
velocity dispersion of $\sigma_{\rm obs,int}$ = 0.06 km~s$^{-1}$ 
after  removing the smearing effect due to the channel width 
using  $\sigma_{\rm obs,int} = \sqrt{ \sigma_{\rm obs}^{2} - 
\sigma_{\rm ch}^{2} }$, where $\sigma_{\rm ch} = 
\triangle_{\rm ch}/(2\sqrt{2\; \rm ln2})$ = 0.089 km~s$^{-1}$ 
is the channel width. 
For the extremely narrow line emission, a higher spectral 
resolution data is required to accurately characterize the actual 
intrinsic velocity dispersion. Considering the 
$\langle \sigma_{\rm obs} \rangle$ is 1.2 times larger than the 
$\sigma_{\rm ch}$, the aforementioned derived 
$\sigma_{\rm obs,int}$ can be regarded as a reasonable 
approximation for the following analysis 
\citep[see Appendix C in][]{2020ApJ...896..110L}.  
Assuming the observed linewidths of the NH$_{2}$D line are caused 
only by thermal broadening, the gas temperature is found to be 
about 7~K, $\sigma_{\rm th,NH_{2}D}$(7~K)\footnote{
The molecular thermal velocity dispersion can be calculated by 
$\sigma_{\rm th}\;=\; \sqrt{(k_{\rm B}T)/(\mu m_{\rm H})} \;=\; 9.08\times 10^{-2} \rm \;km\; s^{-1} (\frac{T}{K})^{0.5}\; \mu^{-0.5}$, where $\mu \;=\; m/m_{\rm H}$ 
is the molecular weight, $m$ is the molecular mass and $m_{\rm H}$ 
is the proton mass.} = $\sigma_{\rm obs,int}$ =  0.06 km~s$^{-1}$. 
This indicates that the gas kinematics is dominated by thermal motions 
with a kinetic temperature that is 7 K at most. 
In addition to M1, there are 
four NH$_{2}$D cores (M9, M12, M13 and M15) that also show 
extremely small velocity dispersions comparable to the NH$_{2}$D 
thermal velocity dispersion in the temperature range of 7 -- 10 K.  
The velocity dispersions of remaining cores are relatively higher 
than the NH$_{2}$D thermal velocity dispersion at  a 
temperature of 10 K ($\sigma_{\rm th, NH_{2}D} \sim$ 0.07 \kms), 
while they are much smaller than the sound speed 
(Figure~\ref{fig:comb}). This indicates that the gas kinematics are   
dominated by thermal motions toward these NH$_{2}$D cores.

Figure~\ref{fig:sigma} shows the annularly averaged observed velocity 
dispersion as a function of the radial distance  ($R_{\rm dist}$) from 
the center of M1. The observed velocity dispersion appears to increase 
with increasing $R_{\rm dist}$ (see Figure~\ref{fig:sigma}), which 
prevails in pre-stellar and starless cores 
\citep{2005ApJ...619..379C,2019ApJ...872..207A}. The increase of 
$\sigma_{\rm obs}$ along R$_{\rm dist}$ is also detected in the 
other 9 cores (M4, M6, M7, M8, M9, M10, M13, M15, M17), 
although the annularly averaged $\sigma_{\rm obs}$ starts to suffer 
from low S/N toward the outer edges of the cores. 
This increasing trend of  $\sigma_{\rm obs}$ along R$_{\rm dist}$ 
indicates that the turbulence is significantly dissipated toward the 
center of the core.  The results indicate that the turbulence 
dissipation is commonly seen toward these NH$_{2}$D cores. 
Alternatively, in the larger-scale collapse scenario, if the pre-stellar 
cores undergo accretion, the infall speed decreases towards the 
center, and thus the line width will be smaller for more central 
regions \citep{2021MNRAS.tmp..412G}.  Although no clear infall 
signatures are detected in M1 using coarser spectral resolution 
(3.0 -- 3.3 \kms) optically thick lines (e.g., HCO$^{+}$ 1-0, HCN 
1-0 and CS 2-1), this possibility cannot be ruled out and should 
be explored using higher spectral resolution data and other 
tracers \citep[e.g., HCN 5-4/4-3/3-2 and HCO$^{+}$ 5-4/3-2;
][]{2014MNRAS.444..874C}.

Overall, the observed small velocity dispersions and their spatial 
distributions resemble those observed in the well known pre-stellar 
cores in Ophiuchus/H-MM1 and L\,1544 
\citep{2007A&A...470..221C,2017A&A...600A..61H,2018ApJ...855..112P}.
This suggests that the identified NH$_{2}$D cores may have 
not been affected by the star formation activities yet.

\subsection{Gas Masses and Dynamical States} 
\label{sec:mass}
Assuming a typical temperature ($T_{\rm dust}$ = 10 K) of pre-stellar 
cores (see also $\S$~\ref{sec:XNH2D}), the gas masses ($M_{\rm gas}$) 
of the NH$_{2}$D cores are estimated from continuum emission in 
two different approaches depending on their intensities (see 
Appendix~\ref{app:mass} for detailed derivation of gas mass).   
First, the continuum emission is used to compute the gas mass 
if the continuum peak emission is higher than $3\sigma$. 
Second, a 3$\sigma$ mass sensitivity of 0.13 \Mo\ is used as an upper 
limit for the core mass if the continuum peak emission is $\leqslant 3\sigma$. 
The derived cores masses are between $<$ 0.13 and 0.87 \Mo, with a mean 
value of  0.45 \Mo\ (Table~\ref{tab:nh2d}).  Following the same procedure,  
the core-averaged H$_{2}$ column densities ($N_{\rm H_{2}}$) are derived 
to be between $<$ 1.1~$\times$~10$^{22}$ and 
3.0~$\times$~10$^{22}$  cm$^{-2}$ (Appendix~\ref{app:mass}), 
with a mean value of 2.0~$\times$~10$^{22}$ cm$^{-2}$.


In order to evaluate the gravitationally bound of the 
NH$_{2}$D cores, we estimated 
the virial masses ($M_{\rm vir} $) that range from 0.43 to 2.54 \Mo\ 
(see Appendix~\ref{app:mass} for derivation of virial mass).  
The derived virial parameters, 
$\alpha_{\rm vir}$ = $M_{\rm vir}/M_{\rm gas}$,  
are between 0.93 and 6.03. Nine out of the 17 NH$_{2}$D cores are 
gravitationally bound with virial parameters $<$ 2, and the remaining 
eight NH$_{2}$D cores are gravitationally unbound with 
$\alpha_{\rm vir} >$ 2  if external pressure is ignored 
 \citep{2013ApJ...779..185K}.

\subsection{NH$_{2}$D Abundances and NH$_{3}$ Deuterium Fractionation} 
\label{sec:XNH2D}
Assuming a constant temperature of 10 K and the ortho-to-para 
ratio of 3 \citep{2017A&A...600A..61H}, the estimated NH$_{2}$D 
column densities ($N(\rm NH_{2}D)$) range from 7.3~$\times$~10$^{12}$~cm$^{-2}$ 
to 3.4~$\times$~10$^{15}$~cm$^{-2}$, with a mean value of 
5.2~$\times$~10$^{14}$~cm$^{-2}$ (see Figure~\ref{fig:comb}). 
The derived abundance 
$X(\rm NH_{2}D)$ = $N(\rm NH_{2}D)$/$N(\rm H_{2})$  toward 
the NH$_{2}$D cores range from 1.8~$\times$~10$^{-9}$ to 
2.0~$\times$~10$^{-7}$, with a mean value of 
2.5~$\times$~10$^{-8}$.

The JVLA NH$_{3}$ observations only cover the central region of 
NGC\,6334S (see Figure~\ref{fig:cont}). There are ten NH$_{2}$D 
cores located in the field of the NH$_{3}$ observations.  Limited by 
the relatively low S/N, both NH$_{3}$ (1,1) and (2,2) transitions are 
detected in M4 and M10 (S/N $\sim$ 4--5), while only NH$_{3}$ (1,1) 
is marginally detected in M13 and undetected in the other seven 
NH$_{2}$D cores. Using \texttt{PySpecKit} we fit the core-averaged 
spectrum, and we derive a kinetic temperature of 
$T_{\rm k}$ = 12.7 $\pm$ 3.1 K and a column density of 
$N_{\rm NH_{3}}$ =  $(4.4\, \pm \,2.5) \times 10^{14}$~cm$^{-2}$  
for M4, and the derived parameters are $T_{\rm k}$ = 12.1 $\pm$ 3.2 K 
and $N_{\rm NH_{3}}$ = $(8.9 \pm 5.0) \times 10^{14}$ cm$^{-2}$ 
for M10, assuming the ortho-to-para ratio of 1 for NH$_{3}$ 
\citep{2017A&A...600A..61H}.  
The column density ratio NH$_{2}$D-to-NH$_{3}$ 
($D_{\rm NH_{3}}$, also known as NH$_{3}$ deuterium fractionation) 
is found to be 0.25 and 0.11 for M4 and M10, respectively. 
A 3$\sigma$ integrated intensity is used to estimate the upper 
limit column density  (3.8 $\times$ 10$^{14}$ cm$^{-2}$) for 
the other eight cores assuming an excitation temperature of 10 K. 
The resulting lower limits $D_{\rm NH_{3}}$ range from 0.11 to 0.39 
(see Table~\ref{tab:nh2d}), which are close to those in the pre-stellar 
core L~1544 \citep[0.5$\pm$0.2;][]{2007A&A...470..221C} and 
Ophiuchus/H-MM1 \citep[0.45$\pm$0.09;][]{2017A&A...600A..61H}.

\section{Discussion} 
\label{sec:dis}
As shown in Figure~\ref{fig:comb}, the derived NH$_{2}$D column 
densities and fractional abundances are higher than  the values of 
Ophiuchus/H-MM1 \citep[$1.1 \,\times\, 10^{14}$~cm$^{-2}$ and 
$2 \,\times\, 10^{-9}$;][]{2017A&A...600A..61H} 
and L1544 \citep[$1.8 \,\times\, 10^{14}$~cm$^{-2}$ and 
$2\, \times \,10^{-10}$;][]{2007A&A...470..221C}. 
The high abundances of NH$_{2}$D could benefit from the 
significant depletion of CO, which occurs faster at cold and 
dense environments  \citep[e.g., starless or pre-stellar 
cores; the detail chemical processes related to the NH$_{2}$D 
formation can be found 
in][]{2015A&A...581A.122S,2017A&A...600A..61H}.

The small velocity dispersion, small Mach number (Table~\ref{tab:nh2d}), 
and positive $\sigma_{\rm obs}$--$R_{\rm dist}$ 
relation indicate that the turbulence is likely dissipated toward these 
NH$_{2}$D cores. In addition, it implies that star formation has not 
yet started; outflows motions would widen the lines as can be seen 
in the continuum cores (Figure~\ref{fig:cont}).  
Therefore, these results in conjunction with the high NH$_{3}$ 
deuterium fractionation, and the lack of embedded YSOs and 70~\um 
emission indicate that the identified NH$_{2}$D cores are still in a 
starless phase. Nine out of 17 starless cores are gravitationally bound 
with $\alpha_{\rm vir} <$ 2, therefore are identified as excellent 
pre-stellar core candidates. 
The dust temperature may drop to $\sim$5.5 K toward the core center 
\citep[e.g., L1544;][]{2007A&A...470..221C}.  The derived gas mass 
can increase by a factor of 2.25 if a temperature of 5.5 K is assumed, 
so that the resulting virial parameter will decrease by a factor of 2.25. 
The levels of the missing flux are unknown due to the lack of single dish 
data. However, the estimated masses and radius of cores may not 
significantly affected by the missing flux, because the observations are 
carried out with the most compact 12~m array configuration (C40-1) 
and the MRS is $\sim$25\arcsec\ that is much larger than the cores sizes.  
Therefore, the estimated virial parameters are not seriously affected 
by the missing flux. 
In addition, the ambient pressure from the parental clumps and 
filaments could also provide additional confinement to make these 
NH$_{2}$D cores bound \citep[e.g.,][]{2019ApJ...877...93C,
2020ApJ...896..110L}. 
Therefore, we cannot completely rule out the possibility that the other 
8 starless cores could also be  pre-stellar core candidates.

The estimated masses of NH$_{2}$D cores range from 0.13 to 
0.87 \Mo, suggesting that they are in the low-mass core regime. 
The majority (15) of NH$_{2}$D cores are associated with filamentary 
structures revealed by the H$^{13}$CO$^{+}$ line emission 
(Li et al. 2021, in preparation).  Therefore, the NH$_{2}$D cores 
have the potential to accrete a significant amount of material from the 
parental clump via filamentary accretion. 
The masses of continuum cores are found to range from low- to 
high-mass regime \citep[0.17 -- 14.03 \Mo;][]{2020ApJ...896..110L}. 
These results suggest that cores with a large range of masses are 
simultaneously forming in this cluster-forming cloud. 
The massive cores are found toward the center of cloud 
\citep[see][]{2020ApJ...896..110L}, while the low-mass cores are 
wide spread throughout the region. 
The evolutionary stages vary significantly over the sample 
of NH$_{2}$D cores, continuum cores, and YSOs (Class~\1 and \2). 
This indicates that there is a temporally continuous formation of low- 
to high-mass cores.

In summary,  NGC\,6334S is forming a stellar protocluster, its 
embedded cores show a wide diversity in masses (0.13 -- 14.03 \Mo) 
and evolutionary stages (pre-stellar -- Class~\2 phases), 
and the embedded cores are expected to grow in mass by 
gas accretion from parental clump via filaments. This seems  to be 
consistent with the competitive accretion massive star formation 
scenario \citep{2006MNRAS.370..488B}, in which the dense cores 
located near the center of the gravitational potential continue 
accreting material to form massive stars.

\section{Summary}
\label{summary}
We present ALMA and JVLA high spatial resolution observations 
toward the massive IRDC NGC\,6334S. We have identified 17 
low-mass starless core candidates that show bright NH$_{2}$D 
emission, but no YSO counterparts and no signs of protostellar 
activity, although we cannot completely rule out the possibility of  
very weak thus undetected outflows or very faint embedded 
low-mass protostars. 
These candidates show almost-thermal velocity 
dispersions (down to $\sigma_{\rm tot} \sim$ 0.06 \kms), 
high NH$_{2}$D abundances (up to $\sim$10$^{-7}$), high 
NH$_{3}$ deuterium fractionations (up to $>$ 0.39), and are 
completely dark in the infrared wavelengths from 3.6 up to 
70~\um. 
In addition, the $\sigma_{\rm obs}$ appear to decrease toward 
the center of cores in 10 of them. These results suggest that  
turbulence has significantly dissipated and the NH$_{2}$D 
abundance has significantly enhanced toward these cores. 
The gas kinematics resemble the well known pre-stellar cores, 
e.g., Ophiuchus/H-MM1 and L\,1544, but with  one to two orders 
of magnitude greater NH$_{2}$D abundance 
\citep{2007A&A...470..221C,2017A&A...600A..61H}. 
Nine out of the 17 NH$_{2}$D cores are gravitationally bound, 
therefore are identified as pre-stellar core candidates.  
Therefore, the NH$_{2}$D line could be a powerful tracer 
to reveal the starless and pre-stellar cores that do not show 
significant dust continuum emission. This is the first detection  
of a cluster of low-mass starless and pre-stellar core candidates in a 
massive star cluster-forming cloud. Low- to high-mass 
cores are simultaneously forming in this cluster-forming cloud.

\acknowledgments
We thank the anonymous referee for constructive comments that 
helped improve this paper. 
This work was partially supported by National Natural Science Foundation 
of China (NSFC)(Grant Nos. 11988101, 11911530226). 
C.W.L. is supported by the Basic Science Research Program through 
the National Research Foundation of Korea (NRF) funded by the 
Ministry of Education, Science and Technology (NRF-2019R1A2C1010851).
H.B. acknowledges support from the European Research Council 
under the European Community's Horizon 2020 framework program 
(2014-2020) via the ERC Consolidator Grant `From Cloud to Star 
Formation (CSF)' (project number 648505). H.B. also acknowledges 
funding from the Deutsche Forschungsgemeinschaft (DFG) via the 
Collaborative Research Center (SFB 881) `The Milky  Way System' 
(subproject B1). 
I.J.-S. has received partial support from the Spanish State Research 
Agency (AEI; project number PID2019-105552RB-C41).
K.Q. is partially supported by National Key R\&D Program of China 
No. 2017YFA0402600, and acknowledges the National Natural 
Science Foundation of China (NSFC) grant U1731237.
A.P. acknowledges financial support from the UNAM-PAPIIT IN111421 
grant, the Sistema Nacional de Investigadores of CONACyT, and from 
the CONACyT project number 86372 of the `Ciencia de Frontera 2019’ 
program, entitled `Citlalc\'oatl: A multiscale study at the new frontier of 
the formation and early evolution of stars and planetary systems’, M\'exico.
J.M.G. acknowledges the support of the grant AYA2017-84390-C2-R 
(AEI/FEDER, EU).
K.W. acknowledges support by the National Key Research and 
Development Program of China (2017YFA0402702, 2019YFA0405100), 
the National Science Foundation of China (11973013, 11721303), NSFC 
grant 11629302, and 
the starting grant at the Kavli Institute for Astronomy and Astrophysics, 
Peking University (7101502287).
H.B.L. is supported by the Ministry of Science and Technology (MoST) 
of Taiwan (Grant Nos. 108-2112-M-001-002-MY3). 
This paper makes use of the following ALMA data: 
ADS/JAO.ALMA\#2016.1.00951.S. ALMA is a partnership of ESO 
(representing its member states), NSF (USA) and NINS (Japan), 
together with NRC (Canada), MOST and ASIAA (Taiwan), and KASI 
(Republic of Korea), in cooperation with the Republic of Chile. The 
Joint ALMA Observatory is operated by ESO, AUI/NRAO and NAOJ.

\newpage
\vspace{5mm}
\facilities{ALMA, JVLA, \texttt{Herschel}.}

\software{ CASA \citep{2007ASPC..376..127M}, 
APLpy \citep{2012ascl.soft08017R}, 
Astropy \citep{2013A&A...558A..33A}, 
Matplotlib \citep{4160265}, 
PySpecKit \citep{2011ascl.soft09001G}.
}


\newpage
\begin{deluxetable*}{cccccccccccccccccccc}
\setlength{\tabcolsep}{0.42mm}{
\tabletypesize{\scriptsize}
\rotate
\tablecolumns{20}
\tablewidth{0pc}
\tablecaption{Physical parameters of the NH$_{2}$D cores. \label{tab:nh2d}}
\tablehead{
\colhead{Core ID}  &\colhead{R.A.}  &\colhead{Decl.}	
&\colhead{$\rm Maj \times Min$}	&\colhead{P.A.}  	&\colhead{$R$}	
&\colhead{$\sigma_{\rm obs}$}	&\colhead{$v_{\rm LSR}$} 	&\colhead{$\mathcal{M}$}	
&\colhead{$W_{\rm NH_{2}D}^{\rm peak}$}
&\colhead{$W_{\rm NH_{2}D}$} &\colhead{$S_{\rm \nu}^{\rm peak}$}&\colhead{$S_{\rm \nu}$}
&\colhead{$M_{\rm gas}$} 	&\colhead{$M_{\rm vir}$} 	&\colhead{$\alpha_{\rm vir}$}
&\colhead{$N_{\rm NH_{2}D}$}	&\colhead{$N_{\rm H_{2}}$}
&\colhead{$X(\rm  NH_{2}D)$} &\colhead{$D_{\rm NH_{3}}$} \\
   	&\colhead{(J2000)}  &\colhead{(J2000)}	
&\colhead{(\arcsec$\times$\arcsec)}	&\colhead{(deg)}  &\colhead{(pc)}
&\colhead{(km/s)}  &\colhead{(km/s)}  &
&\colhead{(Jy/beam km/s)}
&\colhead{(Jy/beam km/s)}  &\colhead{(mJy/beam)} 	&\colhead{(mJy)}
&\colhead{($M_{\odot}$)} 	&\colhead{($M_{\odot}$)} 	&
&\colhead{(cm$^{-2}$)}	&\colhead{(cm$^{-2}$)}
& &   	 
}
\decimalcolnumbers
\startdata
M1 & 17:19:09.38 & -36:03:37.96 & 7.1$\times$2.8 & 107.6 & 0.028 & 0.11 & -3.59 & $<$0.68 & 0.69 & 140.95 & 0.174 & 0.294 & 0.43 & $<$0.68 & $<$1.57 & 2.2E+15 & 1.8E+22 & 1.1E-07 & …	\\	
M2 & 17:19:04.28 & -36:06:57.17 & 9.3$\times$4.3 & 126.6 & 0.040 & 0.30 & -1.36 & 2.41 & 0.40 & 136.13 & 0.186 & 0.412 & 0.61 & 2.54 & 4.18 & 6.1E+14 & 1.6E+22 & 3.2E-08 & $>$0.39	\\	
M3 & 17:19:05.61 & -36:11:01.29 & 7.5$\times$3.9 & 75.4 & 0.034 & 0.20 & -2.95 & 1.38 & 0.41 & 104.66 & 0.292 & 0.462 & 0.68 & 1.25 & 1.84 & 7.0E+14 & 2.4E+22 & 2.8E-08 & …	\\	
M4 & 17:19:11.12 & -36:05:41.37 & 9.4$\times$3.2 & 79.5 & 0.034 & 0.17 & -5.07 & 1.10 & 0.31 & 82.90 & 0.214 & 0.589 & 0.87 & 1.00 & 1.15 & 4.8E+14 & 2.8E+22 & 1.8E-08 & 0.25	\\	
M5 & 17:19:07.39 & -36:10:24.52 & 7.5$\times$2.1 & 72.1 & 0.025 & 0.15 & -2.47 & 0.84 & 0.44 & 82.36 & 0.180 & 0.258 & 0.38 & 0.63 & 1.65 & 9.3E+14 & 1.8E+22 & 4.5E-08 & …	\\	
M6 & 17:19:12.04 & -36:06:55.76 & 7.3$\times$3.9 & 57.3 & 0.034 & 0.18 & -2.80 & 1.13 & 0.28 & 69.97 & 0.230 & 0.458 & 0.68 & 1.08 & 1.60 & 4.0E+14 & 2.4E+22 & 1.7E-08 & $>$0.25	\\	
M7 & 17:19:09.91  & -36:09:05.62 & 6.6$\times$5.2 & 21.1 & 0.037 & 0.23 & -4.17 & 1.69 & 0.23 & 65.62 & 0.206 & 0.447 & 0.66 & 1.83 & 2.77 & 1.8E+14 & 2.2E+22 & 8.0E-09 & …	\\	
M8 & 17:19:10.30 & -36:09:09.49 & 5.8$\times$4.4 & 122.3 & 0.032 & 0.19 & -4.22 & 1.32 & 0.24 & 56.76 & 0.237 & 0.450 & 0.66 & 1.27 & 1.91 & 3.0E+14 & 2.5E+22 & 1.2E-08 & …	\\	
M9 & 17:19:11.40  & -36:08:44.12 & 4.9$\times$3.9 & 78.0 & 0.028 & 0.12 & -1.68 & 0.34 & 0.24 & 45.38 & 0.158 & 0.201 & 0.30 & 0.73 & 2.47 & 2.7E+14 & 1.4E+22 & 1.4E-08 & 0.22	\\	
M10 & 17:19:05.12 & -36:06:45.00 & 4.2$\times$3.7 & 9.9 & 0.025 & 0.15 & -3.47 & 0.78 & 0.26 & 45.19 & 0.310 & 0.366 & 0.54 & 0.79 & 1.46 & 6.9E+14 & 2.8E+22 & 2.5E-08 & $>$0.11	\\	
M11 & 17:18:58.66 & -36:07:04.76 & 4.1$\times$2.2 & 48.5 & 0.019 & 0.11 & -2.01 & $<$0.73 & 0.33 & 44.30 & … & … & $<$0.13 & $<$0.51 & 3.82 & 7.5E+14 & $<$1.5E+22 & $>$7.1E-08 & $>$0.32	\\	
M12 & 17:19:06.86 & -36:06:22.65 & 6.8$\times$3.6 & 124.3 & 0.031 & 0.13 & -3.51 & 0.39 & 0.18 & 41.11 & 0.244 & 0.538 & 0.79 & 0.74 & 0.93 & 2.5E+14 & 3.0E+22 & 8.9E-09 & $>$0.16	\\	
M13 & 17:19:07.39 & -36:05:55.13 & 6.9$\times$3.7 & 116.2 & 0.032 & 0.10 & -5.36 & $<$0.57 & 0.15 & 36.21 & … & … & $<$0.13 & $<$0.80 & 6.03 & 1.3E+14 & $<$1.5E+22 & $>$7.8E-09 & $>$0.14	\\	
M14 & 17:19:04.55 & -36:05:31.47 & 3.7$\times$2.2 & 29.7 & 0.018 & 0.16 & -4.20 & 0.93 & 0.26 & 33.59 & … & … & $<$0.13 & 0.54 & $>$4.04 & 2.8E+14 & $<$1.5E+22 & $>$1.2E-08 & …	\\	
M15 & 17:19:08.88 & -36:08:01.55 & 5.3$\times$2.7 & 78.2 & 0.024 & 0.11 & -4.26 & $<$0.70 & 0.19 & 29.68 & … & … & $<$0.13 & $<$0.62 & 4.64 & 2.0E+14 & $<$1.5E+22 & $>$2.0E-08 & $>$0.17	\\	
M16 & 17:19:07.21 & -36:08:16.60 & 6.4$\times$2.3 & 125.7 & 0.024 & 0.15 & -4.18 & 0.81 & 0.16 & 29.59 & 0.160 & 0.168 & 0.25 & 0.60 & 2.44 & 1.6E+14 & 1.2E+22 & 1.1E-08 & $>$0.14	\\	
M17 & 17:19:07.78 & -36:10:07.54 & 4.8$\times$1.8 & 57.5 & 0.019 & 0.13 & -2.52 & 0.49 & 0.13 & 18.00 & 0.158 & 0.170 & 0.25 & 0.43 & 1.70 & 1.6E+14 & 1.6E+22 & 7.2E-09 & …	\\	\hline
Mean &  &  & 	 &  & 0.028 & 0.16 &  & 0.96 & 0.29 & 62.50 & 0.211 & 0.370 & 0.45 & 0.94 & 2.60 & 5.1E+14 & 2.0E+22 & 2.6E-08 & 0.22	\\	\hline
Median &  &  & 	 &  & 0.028 & 0.15 &  & 0.81 & 0.26 & 45.38 & 0.206 & 0.412 & 0.43 & 0.74 & 1.91 & 3.0E+14 & 1.8E+22 & 1.7E-08 & 0.20	\\	\hline
Minimum &  &  & 	 &  & 0.018 & 0.10 &  & 0.34 & 0.13 & 18.00 & 0.158 & 0.168 & 0.13 & 0.43 & 0.93 & 1.3E+14 & 1.2E+22 & 7.2E-09 & 0.11	\\	\hline
Maximum &  &  & 	 &  & 0.040 & 0.30 &  & 2.41 & 0.69 & 140.95 & 0.310 & 0.589 & 0.87 & 2.54 & 6.03 & 2.2E+15 & 3.0E+22 & 1.1E-07 & 0.39	\\	
\enddata
\tablenotetext{}{Notes. 
For M1, M11, M13, and M15, $\sigma_{\rm nt,NH_{2}D}$ = $\sigma_{\rm obs}$ 
was used since the $\sigma_{\rm obs,int} \leqslant \sigma_{\rm nt,NH_{2}D}(\rm 10 K)$.
(4)-(5) beam-deconvolved size. 
(6) beam-deconvolved effective radius. 
(7)-(8) $\sigma_{\rm obs}$ and $v_{\rm LSR}$ are derived by the 
core-averaged spectrum. 
(9) Mach number $\mathcal{M}$ = $\sqrt{3} \sigma_{\rm nt,NH_{2}D}/c_{\rm s}$. 
(10) the peak value of NH$_{2}$D velocity-integrated image. 
(11) the integrated flux density of the NH$_{2}$D line emission. 
(12)-(13) the peak and integrated intensity of continuum emission. 
(14) gas mass estimated by continuum emission. 
(15) virial mass. 
(16) virial parameter. 
(17) mean NH$_{2}$D column density. 
(18) mean  H$_{2}$ column density. 
(19) mean NH$_{2}$D abundance fraction. 
(20) the NH$_{3}$ deuterium fractionation $D_{\rm NH_{3}}$. 
} 
}
\end{deluxetable*}

\bibliography{prestellar}{}
\bibliographystyle{aasjournal}

%
\appendix
 
\section{NH$_{2}$D core identification}
\label{app:identification}
The \texttt{astrodendro}\footnote{ http://dendrograms.org/} algorithm 
was used to preselect the compact NH$_{2}$D emission structures 
\citep[i.e., the leaves in the terminology of 
\texttt{astrodendro};][]{2008ApJ...679.1338R} from the 
NH$_{2}$D velocity-integrated intensity image (integrated between 
-11.5 and 5.6 \kms). 
The following parameters are used in the computation of 
\texttt{astrodendro}: the minimum pixel value is 3$\sigma$ noise level 
(1$\sigma \sim$ 0.02 Jy beam$^{-1}$ km s$^{-1}$ for $W_{\rm NH_{2}D}$); 
the peak value is $\geqslant$ 6$\sigma$; 
the minimum difference in the peak intensity between neighboring 
compact structures is  1$\sigma$; and the minimum number of pixels 
required for a structure to be considered an independent entity is 40 that 
is approximately the synthesized beam area. 
From the identified leaves, we further search for the compact structures 
that are no associated with continuum cores, YSOs, and showing a 
single peak and no extreme elongations (major axis smaller than 3 times 
minor axis).  
The physical parameters (size, flux density, peak intensity) of these 
selected compact structures are extracted from the NH$_{2}$D 
velocity-integrated intensity image using the  \texttt{CASA-imfit} task 
(see Table~\ref{tab:nh2d}). 
We stress that this identification is likely to be incomplete. Potential 
NH$_{2}$D compact structures could have been missed if they 
cannot be distinguished from the clumpy or filamentary structures.
Four NH$_{2}$D cores (M11, M13, M14, M15) are not associated with 
continuum emission at all 
($< 3\sigma$, 1$\sigma \sim$ 30 $\mu$Jy~beam$^{-1}$ 
for continuum image), and the rest of the NH$_{2}$D cores are partially 
associated with weak continuum emission ($3\sigma - 7\sigma$), 
except for M10 that has about 10\% region at the edge of the core 
associated with $8\sigma - 11\sigma$ continuum emission 
(Figure~\ref{fig:cores}).

\section{Gas mass and Virial mass}
\label{app:mass}
With the observed 3~mm continuum fluxes, we estimated 
the gas mass ($M_{\rm gas}$) of identified cores assuming 
optically thin modified black body emission in the Rayleigh-Jeans 
limit, following 
\begin{equation}
\label{dust_mass}
M_{\rm gas}=\eta \frac{S_{\nu} d^{2}}{B_{\nu}(\rm T)\, \kappa_{\nu}},
\end{equation}
where $\eta$ = 100 is the assumed gas-to-dust mass ratio, 
$d$ is the source distance, 
$S_{\nu}$ is the continuum flux at frequency 98.5 GHz, 
$B_{\nu}(\rm T)$ is the Planck function at  temperature $T$, 
and $\kappa_{\nu}$ is the dust opacity at frequency $\nu$. 
We adopt $\kappa_{\rm 98.5\; GHz}$ = 0.235 cm$^{2}$g$^{-1}$, 
assuming  $\kappa_{\nu}$ = 10(${\nu}$/1.2 THz)$^{\beta}$ 
cm$^{2}$g$^{-1}$ and $\beta$ = 1.5 for all of dense cores  
\citep{1983QJRAS..24..267H}. 

The column density ($N_{\rm H_{2}}$) is computed with 
\begin{equation}
\label{dust_mass}
N_{\rm H_{2}}=\eta \frac{S_{\nu}^{\rm beam}}{B_{\nu}({\rm T})\, \kappa_{\nu} \, \Omega \, \mu \, m_{\rm H}},
\end{equation}
where $S_{\nu}^{\rm beam}$ is the continuum flux density, 
$\Omega$ is the beam solid angle, $\mu$ = 2.8 is the mean 
molecular weight of the interstellar medium 
\citep{2008A&A...487..993K}, and $m_{\rm H}$ is the proton mass.

The virial mass is estimated using:
\begin{equation}
\label{equ:Mvir}
M_{\rm vir}=\frac{5}{a \beta} \frac{\sigma_{\rm tot}^{2} R}{G} ,
\end{equation}
where $\sigma_{\rm tot}\; =\; \sqrt{\sigma_{\rm nt,NH_{2}D}^{2} + \sigma_{\rm th,\langle m \rangle}^{2}}\; = \; \sqrt{(\sigma_{\rm obs,int}^{2} - \sigma_{\rm th,NH_{2}D}^{2}) + \sigma_{\rm th,\langle m \rangle}^{2}}$ 
is the total velocity dispersion, 
$R$ is the effective radius, $G$ is the gravitational constant, 
parameter $a$ equals to $(1-b/3)/(1-2b/5)$ for a power-law 
density profile $\rho \propto r^{-b}$, and $\beta = (\arcsin e)/e$ 
is the geometry factor \citep[see also][]{2013ApJ...768L...5L,2020ApJ...896..110L}. 
Here $\langle m \rangle$ is 2.37$m_{\rm H}$ \citep{2008A&A...487..993K},  
$\sigma_{\rm obs,int}$ is derived  by the core-averaged spectrum, 
and $b$ is assumed to be 2. 
For M1, M11, M13, and M15, 
$\sigma_{\rm tot}$ = $\sigma_{\rm th,\langle m \rangle}(\rm 10\,K)$ 
was used since the 
$\sigma_{\rm obs,int} \leqslant \sigma_{\rm th,NH_{2}D}(\rm 10\, K)$.

\section{Internal Luminosity}
\label{app:lum}
The internal luminosity of dense object can be estimated by using the 
empirical relationship between internal luminosity ($L_{\rm int}$) and 
observed  70 \um flux \citep{2008ApJS..179..249D}:
\begin{equation}
\label{equ:Mvir}
L_{\rm int} (L_{\odot})=3.3\times10^{8} \left(F_{\rm 70}\, \left(\frac{d}{140 \rm pc}\right)^{2} \right)^{0.94},
\end{equation}
where $F_{\rm 70}$ is the flux at 70 \um in cgs units 
(erg s$^{-1}$ cm$^{-2}$  Hz$^{-1}$) and $d$ is the source distance in pc.  
A 3$\sigma$ sensitivity of $\sim$0.3 Jy from \texttt{Herschel}/PACS 70 \um\ 
observation is used to derived an upper limit ($\sim$1.26 $L_{\odot}$) of 
internal luminosity for the identified NH$_{2}$D cores 
\citep{2017A&A...602A..77T}, with the exception of M3, 
M5, and M17, which are illuminated by the western YSOs rather than 
internal heating. The measured 70 \um\ fluxes of  M3, M5, and M7 are 
about 5 -- 6$\sigma$.


\end{document}